\documentstyle[preprint,prb,aps,epsfig,eqsecnum]{revtex}

\newcommand{\be}{\begin{equation}}
\newcommand{\ee}{\end{equation}}
\newcommand{\bn}{\begin{displaymath}}
\newcommand{\en}{\end{displaymath}}
\newcommand{\bs}{\begin{eqnarray}}
\newcommand{\es}{\end{eqnarray}}

\newcommand{\nn}{\nonumber}
\newcommand{\Pf}{\text{Pf}\;}
\newcommand{\up}{{\mathsf up}}
\newcommand{\ri}{{\mathsf right}}
\newcommand{\dw}{{\mathsf down}}
\newcommand{\lf}{{\mathsf left}}
\newcommand{\sq}{{\Box}}
\newcommand{\bigstar}{\ast}

\begin{document}

\tightenlines
\draft

\title{Finite size corrections for the Ising model on higher genus triangular lattices}

\author{Ruben Costa-Santos
\footnote{e-mail R.A.Costa-Santos@phys.uu.nl}}
\address{Spinoza Institute, Utrecht University, Leuvenlaan 4, 3584 CE Utrecht}

\author{Barry~M.~McCoy
\footnote{e-mail mccoy@insti.physics.sunysb.edu}}
\address{C.N. Yang Institute for Theoretical Physics,
 State University of New York at Stony Brook, NY 11794-3840}

\preprint{YITP-SB-02-54/SPIN-2002-30/ITP-2002-49}

\maketitle

\begin{abstract}

\vspace{-.5cm}

We study the topology dependence of the finite size corrections to the Ising model partition function by considering the model on a triangular lattice embedded on a genus two surface.  At criticality we observe a universal shape dependent correction, expressible in terms of Riemann theta functions, that reproduces the modular invariant partition function of the corresponding conformal field theory. The  period matrix characterizing the  moduli parameters of the limiting Riemann surface is obtained by a numerical study of the lattice continuum limit. The same results are reproduced using a discrete holomorphic structure.

\end{abstract}

\vspace{6cm}

\section{Introduction}\label{section0}

In the context of two dimensional critical phenomena, finite size corrections\cite{cardy} provide a direct connection between the critical behavior of specific lattice models  and the conformal field theory description of the corresponding universality class.
For a two dimensional system without boundaries, of typical length $L$, the partition function is expected from very general arguments\cite{cardy2} to behave, at the critical point and  in the large   $L$ limit, as
\be
      Z(L)\simeq A \  L^h \, \exp{(-f_\infty L^2)} \label{caex}
\ee
where $h$ depends on the system topology and singularities of the metric and $A$ depends on the shape of the system.

In this paper we study the shape dependence of the constant term $A$ for the  Ising model, and clarify its connection with conformal field theory, by considering the model on triangular lattices embedded on a genus two surface.  Finite size corrections on higher genus lattices have been previously studied  on square lattices\cite{me} for both the Ising model and the close-packed dimers. Here we test the universality of the results obtained by considering the triangular lattice case. The dimer model is not considered since it is not critical on triangular lattices\cite{fendley}.

For a lattice embedded on a genus $g$ surface, the Ising model partition function can be expressed, using the Kasteleyn dimer formalism\cite{kast1,kast2,kast3,russ,tesl}, in terms of Pfaffians of $4^g$ adjacency matrices $A_k$
\be
   Z_g(N_i,K_i)= \sum_{k=1}^{4^g} \, \Pf A_k(N_i,K_i) \label{zkast}  .
\ee
where the $N_i$ and $K_i$ are respectively the integer sizes and coupling constants characterizing the lattice.
 We study numerically these Pfaffians in a number of genus $g=2$ triangular lattices and show that they satisfy at the critical point and in the large lattice size ${\cal N}$ limit
\be
    \lim_{{\cal N}\rightarrow \infty}\frac{ \Pf A_k(N_i,K_i)}{ \Pf A_l(N_i,K_i)}= \left|\frac{ \Theta[k](0|\Omega)}{ \Theta[l](0|\Omega)}\right| \label{ad}
\ee
where  the $\Theta[i](0|\Omega)$ are genus two Riemann theta functions, defined in Section \ref{section2}, and $\Omega$ is the $2\times 2$ period matrix.
The shape dependent term in the partition function (\ref{caex}) is found to be, in the large lattice limit,
\be
         A \simeq  \tilde A\  \sum_{k=1}^{16}\left| \Theta[k](0|\Omega)\right| \label{shape}
\ee
where $\tilde A$ is a $k$-independent constant.

This sum over theta functions can be identified with the classical winding part of the partition function of the corresponding conformal field theory\cite{alvarez1,verver1,g2char} on a genus two Riemann surface with a complex structure characterized by the period matrix $\Omega$.

The remarkable similarity between the combinatorial result (\ref{zkast}) and  the sum over $4^g$ fermion boundary conditions\cite{alvarez1} in conformal field theory, together with the explicit results obtained so far for genus one\cite{ferdfish,nashoco2} and two\cite{me} suggest that the shape dependent correction (\ref{shape}) is universal.
By this we mean the following; the expression (\ref{shape}), with a sum over $4^g$ theta function for more general topology, is found at criticality for all lattices independently of the lattice structural details, such as aspect ratios and coupling constants, while these structural details determine the $g\times g$ period matrix $\Omega$.

We are left with the problem of understanding the dependence of the period matrix $\Omega$ on the lattice  detailed structure.
For toroidal lattices this is simply the lattice aspect ratio weighted by the coupling constants along the two lattice directions\cite{ferdfish,nashoco2,brankov,luwu2}.
In the higher genus case we will show that the lattice period matrix $\Omega$ can be evaluated as the continuum limit of a discrete period matrix obtained by solving certain sets of finite difference Poisson equations on the lattice.

The concept of discrete period matrix was first introduced by the authors\cite{me}, in the context of locally square lattices, using a discrete formulation of harmonic and holomorphic differentials. While discrete harmonic differentials are well understood, in the context of combinatorial Hodge theories on lattice cochains\cite{eckmann,dodziuk,dodziuk2}, discrete holomorphy is more elusive, the main  obstacle to a lattice formulation being the discretization of the Hodge operator. Discrete Hodge operators have a long history  in  finite difference  calculations in electrodynamics and other problems of computational physics\cite{books,kotiuga,bourdlala,tarketboss,teix1,hiptmair}. Two dimensional formulations have also been studied\cite{mercat,mercat3} in connection with the Ising criticality and discrete holomorphy.

 In this paper we extend our previous study\cite{me} to triangular lattices. We show that the period matrix for a triangular lattice $G$ can be determined in terms of an effective square lattice $G_\sq$. This correspondence depends explicitly on the  Ising model criticality condition and  provides a geometrical interpretation of this condition.

\begin{figure}[h]
\vskip 2cm
\epsfig{figure=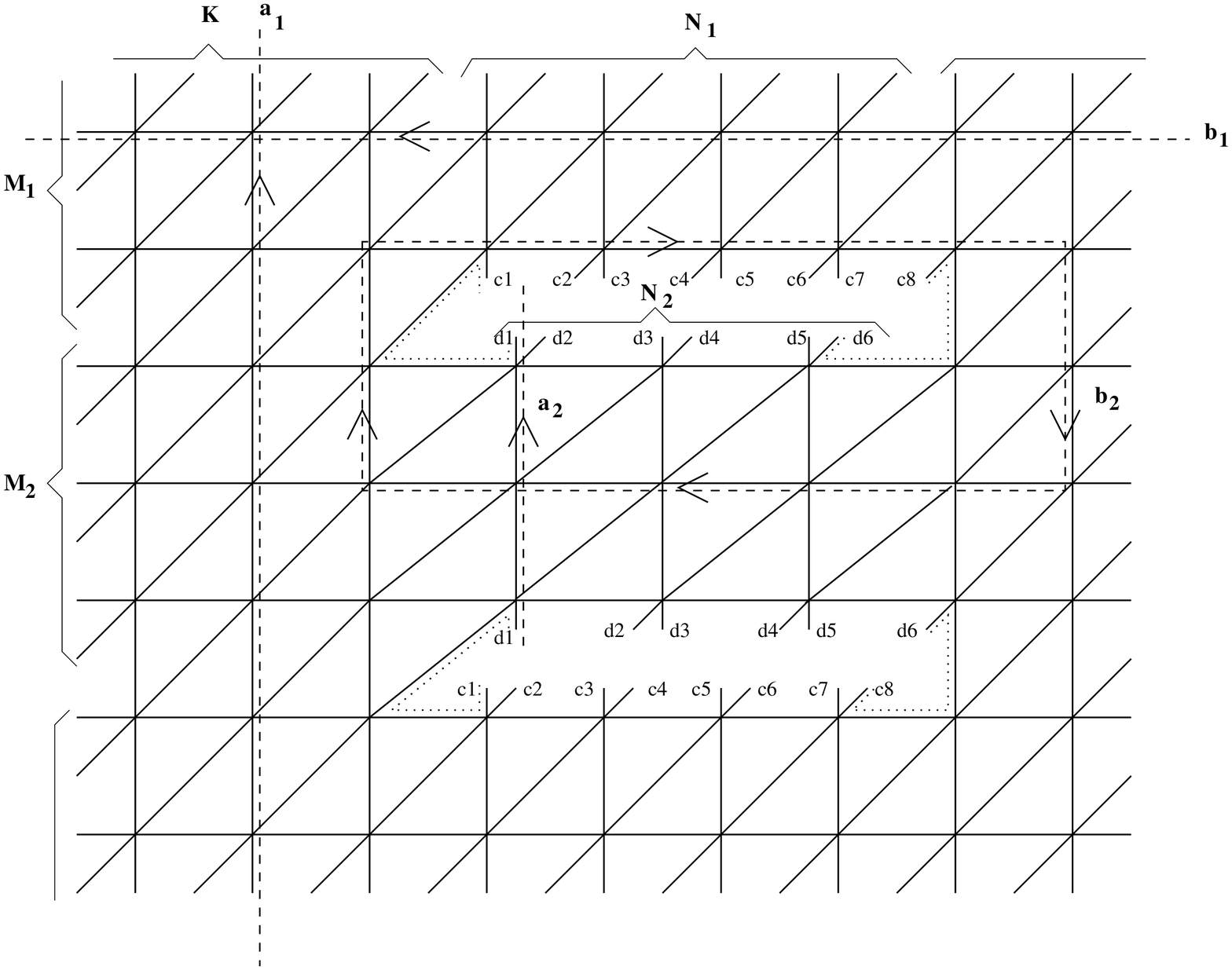,width=15cm}
\vskip 1cm
\caption{The genus two triangular lattice. The exterior boundary has torus-like boundary conditions, the higher genus topology is due to the additional handle obtained by identifying the edges numbered with the c and d numbers. The $a_i,b_i$ cycles form a basis of the first homology group and should be seen as drawn over the lattice edges. All lattice faces are triangular except for two hexagons whose edges are marked with the dotted line.}
\label{fig1}
\end{figure}

The structure of the paper is as follows:

In section \ref{section1}  we discuss the Kasteleyn formalism and define the adjacency matrices in (\ref{zkast}) for the genus two lattice  shown in Fig. \ref{fig1}. In section \ref{section2} we study (\ref{ad}) numerically by evaluating ratios of determinants of adjacency matrices. In section \ref{section3}  we define  discrete harmonic differentials. In section \ref{section4} we define discrete holomorphic differentials and the lattice period matrix and discuss the relation between our formalism and the formalism of reference\cite{mercat3}.  The period matrices of (\ref{shape}) are reproduced using discrete holomorphy in section \ref{section5}.
Finally, our conclusions are presented in section \ref{section6}.

\section{The Kasteleyn formalism} \label{section1}

The partition function of the Ising model on  a lattice, or more
generally on a graph, embedded  without superposition of edges on
an orientable surface of genus $g$ can be expressed  as a sum (\ref{zkast}) 
 over the  Pfaffians of $4^g$ adjacency matrices.
This result, due to  Kasteleyn\cite{kast3,mccoy,russ,tesl},   is
the starting point for the study of the Ising model on higher
genus lattices.

\begin{figure}[h]
\vskip 3cm
\epsfig{figure=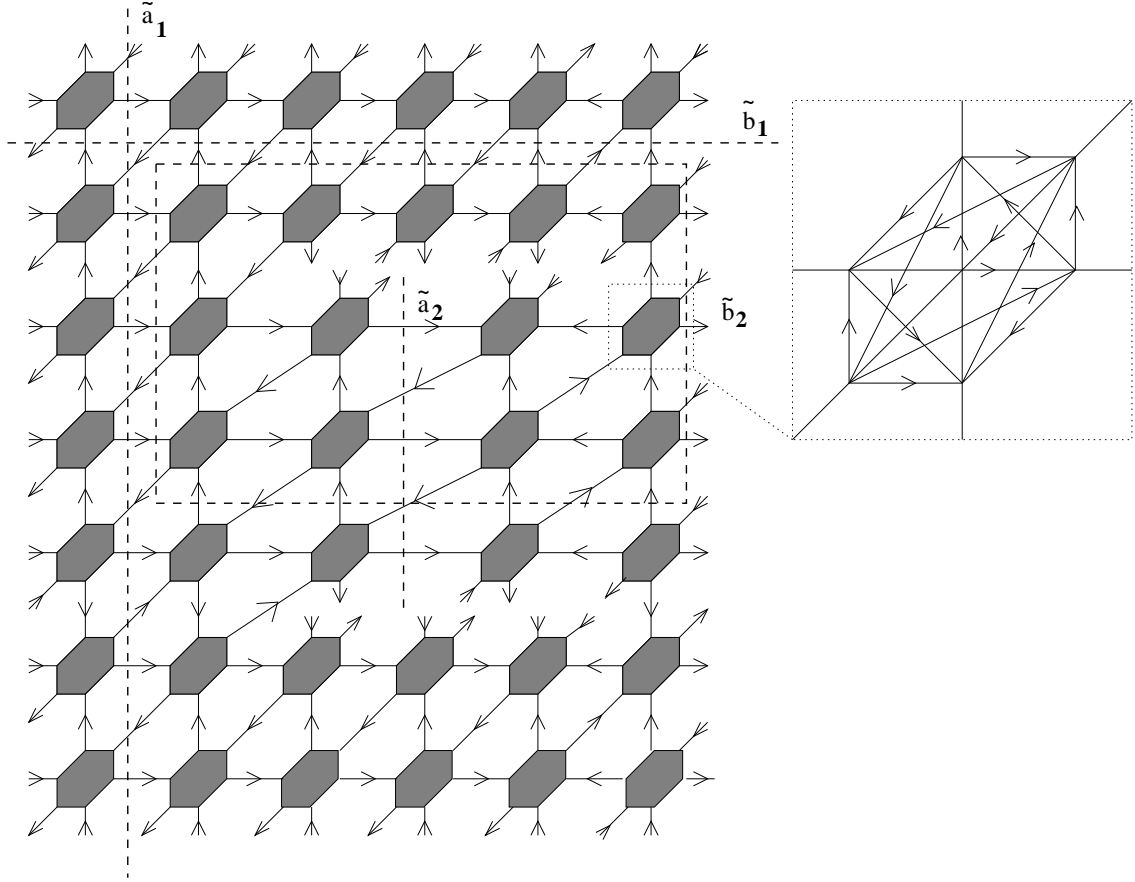,width=15cm}
\vskip 1cm
\caption{The Ising decorated lattice in the $A(0000)$ clockwise odd orientation. The orientation of the hexagonal decoration is shown on the inset. The remaining clockwise odd orientations can be obtained from this one by inverting the orientations of the edges crossed by a given choice of the  $\tilde a_i,\tilde b_i$ cycles.}
\label{fig2}
\end{figure}

In this section we present the main aspects of the Kasteleyn formalism and define the adjacency matrices for the genus two triangular lattice, $G$, shown in  Fig. \ref{fig1}. This  lattice is characterized by five integers sizes $(M_1,M_2,K,N_1,N_2)$ and three coupling constants ($K_h$,$K_v$,$K_d$) distinguishing the horizontal, vertical and diagonal edges.
The boundary conditions are such that the lattice can be drawn without superposition of edges only on a surface of genus two or higher. The cycles of lattice edges $ a_i, b_i$, with $i=1,2$, represent a canonical basis of the first homology group of the embedding surface. The same cycles drawn over the dual lattice edges will be denoted by $\tilde a_i, \tilde b_i$ (see Fig. \ref{fig2}).

The Ising model on the lattice  $G$ can be expressed in terms of a dimer model on a decorated lattice $G'$ using a well known\cite{kast2,steph} one to one correspondence between polygon configurations of the Ising model high temperature expansion in  $G$ and  close-packed dimer configurations in $G'$.  A close-packed dimer configuration is a selection of lattice edges such that every vertex is included once and only once as a boundary of a selected edge.

The decorated lattice $G'$ is obtained by replacing each vertex of the original lattice $G$ with a decoration graph. For a triangular lattice the decoration can be chosen\cite{steph} to be a hexagon with all vertices connected  (see Fig. \ref{fig2}) and the decorated lattice will have $6{\cal N}$ vertices for a original lattice of size ${\cal N}$.

Each edge on the decorated lattice $G'$ is assigned a weight: $w_i=\tanh{\beta K_i }$ for edges of type $i$ inherited from the lattice $G$ and a weight $1$ for the internal edges of the decoration. The Ising partition function on the lattice $G$ can then be expressed as
\be
  Z^{\text{Ising}} = (2 \cosh{K_v}\cosh{K_h}\cosh{K_d})^{{{\cal N}}} \;\sum_{\text{dimer config.}}^{G'}  {w_v}^{n_v} {w_h}^{n_h} {w_d}^{n_d}   \label{dimerpart}
\ee
were the sum runs over all the close-packed dimer configurations of the decorated lattice $G'$ and $n_i$ is the number of edges of type $i$ in a given dimer configuration.

The dimer partition function can be expressed in terms of Pfaffians of the Kasteleyn adjacency matrices.
These are signed adjacency matrices defined in the following way:  label the decorated lattice vertices with an integer from 1 to ${\cal M}$ and choose a lattice edge orientation by assigning to each edge a direction represented by an arrow (see Fig.  \ref{fig2}). The signed adjacency matrix corresponding to this edge orientation is the ${\cal M}\times {\cal M}$ matrix $A_{ij}$ with entries

\be
       A_{ij}= \left\{
              \begin{array}{rl}
              z &  \textstyle{\text{ if there is an arrow from vertex i to vertex j of weight }} z \\
             -z &  \textstyle{\text{ if there is an arrow from vertex j to vertex i of weight }} z\nn \\
              0 & \textstyle{\text{ otherwise}}
               \end{array}   \right.
\ee

The Pfaffian of such an ${\cal M}\times {\cal M}$ anti-symmetric matrix, with ${\cal M}$ even, is defined as
\be
    \Pf  (A) = \frac{1}{2^{{\cal M}/2}({\cal M}/2)!} \sum_p \epsilon_p \;A_{p_{1}p_{2}}A_{p_{3}p_{4}}\cdots A_{p_{{\cal M}-1}p_{{\cal M}}} \label{pff}
\ee
where the sum goes over all the permutations $p$ of the integers from 1 to ${\cal M}$ and $\epsilon_p=\pm 1$ for even and odd permutations respectively.

{}From the definition of the adjacency matrix it is clear that each non zero term in the Pfaffian equals, in absolute value, a term in the dimer partition function, with the relative sign
 depending on the choice of the edge orientation. If we choose an orientation  such  that all the terms in the Pfaffian have the same relative sign, the Pfaffian of the adjacency matrix equals the partition function modulo an overall sign.

Kasteleyn showed in a beautiful tour de force of combinatorics\cite{kast1} that edge orientations with this property exist. These are the edge orientations such that every lattice face has an odd number of clockwise oriented edges. In a genus $g$ lattice there will be  $4^g$ inequivalent such orientations and the dimer partition function is given as a linear combination of their Pfaffians, see references\cite{kast3,russ,tesl} for more details.

For the genus two decorated lattice $G'$  the relevant
sixteen edge orientation classes can be labelled as $A(n_{\tilde
a_1},n_{\tilde b_1},n_{\tilde a_2},n_{\tilde b_2})$ with $n_x=0,1$
for $x= \tilde a_1,\tilde a_2,\tilde b_1,\tilde b_2$.  The
orientation $A(0000)$  is shown in Fig. \ref{fig2}. An orientation
with a certain $n_x=1$ is obtained from the corresponding
orientation with $n_x=0$ by introducing a disorder loop along the
non-trivial cycle $x$, this is, by reversing the orientation of
all the edges crossed by the cycle $x$.

To  make connection with the theta functions characteristic and allow for more compact equations we will also use the alternative notations
\be
   A(n_{\tilde a_1},n_{\tilde b_1},n_{\tilde a_2},n_{\tilde b_2})= A\scriptsize{\left[\begin{array}{cc}   n_{\tilde b_1} & n_{\tilde b_2} \\ n_{\tilde a_1} & n_{\tilde a_2} \end{array} \right]}=A_i \label{orient} \label{defori}
\ee
 with the integer label given by $i={\textstyle 16-8n_{\tilde a_1}- 4n_{\tilde b_1}-2n_{\tilde a_2}-n_{\tilde b_2}}$.
In terms of these adjacency matrices the Ising model partition function on $G$ is given by
\bs
   \begin{array}{rccccccccccccccccl}
Z=\frac{1}{4}(&P_1&-&P_2&-&P_3&-&P_4&-&P_5&+&P_6&+&P_7&+&P_8&-&  \\
     &P_{9}&+&P_{10}&+&P_{11}&+&P_{12}&-&P_{13}&+&P_{14}&+&P_{15}&+&P_{16}&)& \end{array} \label{zzz}
\es
where $P_i=\Pf A_i$ and the adjacency matrices are defined on the lattice $G'$. 

For translational invariant lattices, such as the torus triangular lattice, the Pfaffians of the adjacency matrices can be evaluated in a closed form and the theta function dependence  extracted by a careful asymptotic analysis\cite{ferdfish,nashoco2}
For a genus two lattice as $G'$ such an analytic treatment is not possible and we are forced to resort to numerical evaluations of the Pfaffians.

\section{Ratios of determinants and theta functions}\label{section2}

The Pfaffians of the adjacency matrices defined on the previous section can be numerically evaluated using the fact that $\Pf A\!\!=\!\!\sqrt{\det A}$ for an anti-symmetric matrix A.
We evaluate  these determinant for a sequence of lattices characterized by the integers sizes $(M_1,M_2,K,N_1,N_2)= (m_1,m_2,k,n_1,n_2)\,L$ with increasing $L$ and fixed aspect ratios  $(m_1,m_2,k,n_1,n_2)$.
The Ising model coupling constants were restricted to the ferromagnetic case $K_i\geq 0$ and chosen to satisfy the criticality condition
\be
    \sinh(2  K_h)\sinh(2  K_v)+\sinh(2  K_h)\sinh(2  K_d)+ \sinh(2  K_v)\sinh(2  K_d)=1. \label{critic}
\ee

\begin{figure}[h]
\vskip 3cm
\epsfig{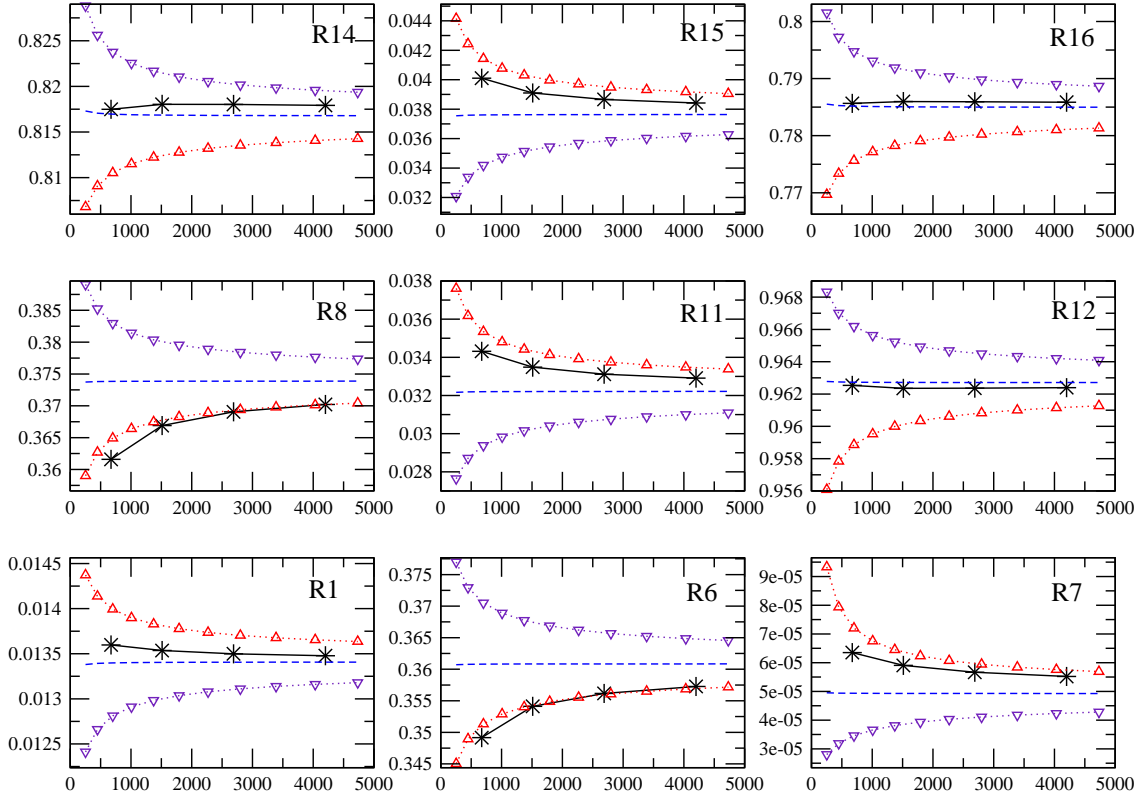}
\vskip 1cm
\caption{Comparison between the nine non-vanishing ratios of determinants of adjacency matrices (full line and star symbols) and ratios of theta functions (dotted and dashed lines and triangle symbols) with period matrices evaluated by the procedure of section \ref{section4}. Ratios are shown as function of the number of lattice vertices $\cal N$, for a lattice with aspect ratio ($m_1$,$m_2$,k,$n_1$,$n_2$)=(4,2,2,2,4)  and couplings $[w_h,w_v,w_d]$=[0.3685,0.1586,0.2821]. The period matrix evaluations with triangle up and triangle down symbols correspond respectively to evaluations done on the direct and on the dual lattice. The dashed line is the theta ratio for the average of the two period matrices.}
\label{fig3}
\end{figure}

\begin{figure}[h]
\vskip 3cm
\epsfig{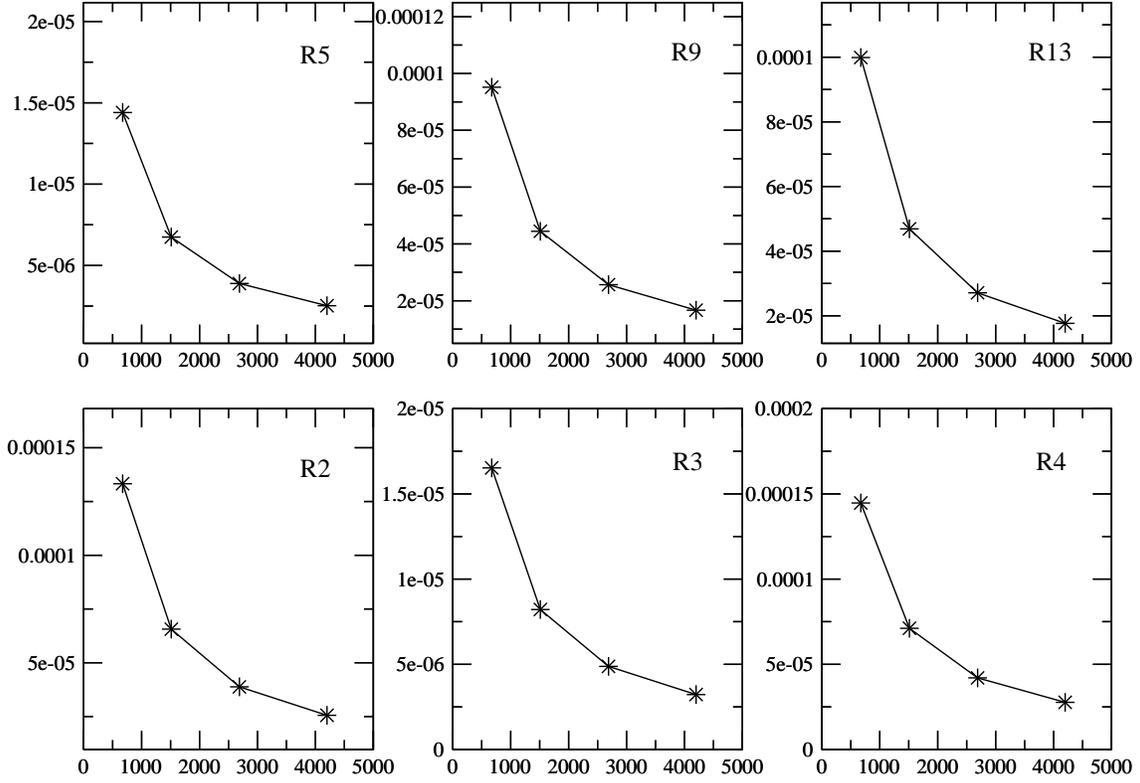}
\vskip 1cm
\caption{The six vanishing ratios of adjacency matrices determinants for the lattice of Fig. \ref{fig3}. These ratios correspond to theta functions of odd half-integer characteristic.}
\label{fig4}
\end{figure}

In order to eliminate the common bulk and algebraic terms, we consider ratios of the sixteen determinants to the largest among them
\be
R_i=\frac{\det{A_i}}{ \det{A_{\text{max}}}} . \label{ratii}
\ee
 An example of the results obtained is shown in Fig. \ref{fig3} and Fig. \ref{fig4}, where the fifteen ratios are shown, for a lattice  with aspect ratios  $(m_1,m_2,k,n_1,n_2)=(4,2,2,2,4)$ and coupling constants  $[w_h,w_v,w_d]=[0.369,0.159,0.282]$. The ratios are shown as function of the number of lattice vertices,  ${\cal N}$.  Six of them, shown in Fig. \ref{fig4},  are found to vanish in the large ${\cal N}$ limit while the remaining nine converge to finite values, Fig. \ref{fig3}.

We found that these  ratios in the large ${\cal N}$ limit can be expressed in terms of ratios of genus two Riemann theta functions. These functions are defined\cite{mumford} by the quickly converging series
\be
   \Theta {\left[\begin{array}{c}  { \mbox{\boldmath $\alpha$}} \\  { \mbox{\boldmath $\beta$}} \end{array} \right]}  \left({\bf z},\Omega \right)
   =\sum_{{\bf n}\in Z^2} \exp{\left[i\pi({\bf n}+ { \mbox{\boldmath $\alpha$}})^T \Omega({\bf n}+{ \mbox{\boldmath $\alpha$}})+2\pi i({\bf n}+{ \mbox{\boldmath $\alpha$}})^T ({\bf z}+{ \mbox{\boldmath $\beta$}})\right]}   \label{theta}
\ee
where {\boldmath $\alpha$}, {\boldmath  $\beta$}, ${\bf  z}$ and ${\bf n}$ are  2-vectors half-integers, complex numbers and integers respectively. The $2\times 2$ period matrix $\Omega$ is a symmetric complex matrix with positive definite imaginary part
\be
     \Omega=\left[\begin{array}{cc} \Omega^r_{11} &\Omega^r_{12}\\\Omega^r_{12}& \Omega^r_{22}\end{array} \right] +i\left[\begin{array}{cc} \Omega^i_{11} &\Omega^i_{12}\\\Omega^i_{12}& \Omega^i_{22}\end{array} \right].
\ee

The extrapolated values of the determinants ratios, in the  $L\rightarrow \infty$  limit, can be expressed  as
\be
  \left. \frac{\det\left( A\scriptsize{\left[\begin{array}{cc} c_1&c_2\\c_3&c_4\end{array} \right]}\right)}
             {\det \left(A\scriptsize{\left[\begin{array}{cc}0 & 0\\ 0 & 0 \end{array} \right]}\right)} \,\right|_{T_c} =
  \left|\frac{\Theta\scriptsize{\left[\begin{array}{cc} c_1/2&c_2/2\\c_3/2&c_4/2  \end{array} \right]}\left(0,\Omega \right)}
             {\Theta\scriptsize{\left[\begin{array}{cc}  0&0\\ 0&0 \end{array} \right]}\left(0,\Omega \right)}\right|^2 \label{resultati}
\ee
for the 16 combinations of $c_i=0,1$, with a remarkable precision from $10^{-4}$ to $10^{-6}$, with a period matrix $\Omega$ determined by a suitable numerical fitting procedure.
Examples  for three different lattices  are shown in  table \ref{table1}. For each lattice the first column shows  the  $L\rightarrow \infty$ extrapolated ratios of determinants (\ref{ratii}) and the second column shows the theta function ratios
\be
     \Theta_{(16-8d_1-4c_1-2d_2-c_2)}(\Omega)=\Theta{\scriptsize \left[\begin{array}{cc} c_1/2& c_2/2\\ d_1/2& d_2/2\end{array} \right]}\left(0,\Omega \right)/\Theta{\scriptsize\left[\begin{array}{cc} 0&0\\ 0&0\end{array} \right]}\left(0,\Omega \right)
\ee
for $c_i,d_i=0,1$,  with the fitted period matrix.

 The period matrices in (\ref{resultati}) are  in general complex  for triangular lattices  unlike what was found for square lattices\cite{me} where the period matrix is purely imaginary.
 Away from criticality, the sixteen determinants converge rapidly to the same value and (\ref{resultati})  does not hold.

\section{Harmonic differentials on the lattice}\label{section3}

 We now turn to the problem of evaluating the period matrix $\Omega$ in (\ref{resultati}) directly from the lattice structure using a discrete formulation of harmonic and holomorphic differentials.

 In continuum Riemann surface theory the period matrix characterizing a given surface is evaluated in two steps, the first step being the determination of the space of harmonic 1-forms on the surface, the second step being the decomposition of this space into holomorphic and anti-holomorphic sub-spaces.
On a lattice the role of p-forms is played by p-cochains, which are linear functionals defined on formal sums of the lattice p-elements: vertices, edges and faces. We will refer to the p-cochains by lattice functions, lattice differentials and lattice volume forms. 

In this section we give the first step towards a definition of the lattice period matrix by defining and evaluating  the harmonic lattice differentials or 1-cochains on the lattice. Discrete harmonicity and the relation between p-cochains and p-forms is well understood from the work of Eckmann\cite{eckmann} and Dodziuk and Patodi\cite{dodziuk,dodziuk2} on combinatorial Hodge theories. 
Here we generalize the treatment done in reference\cite{me} to triangular lattices by showing that harmonic differentials on the triangular lattice are completely determined by their restriction to an effective square lattice.

\subsection{Lattice differentials on a triangular lattice}

A lattice function $f$  is
determined by its value on the lattice vertices $f[n]:
n=1,\ldots,\cal N$ with $n$ being an integer labeling of the
$\cal N$  vertices in the lattice.

\begin{figure}[h]
\vskip 3cm
\epsfig{figure=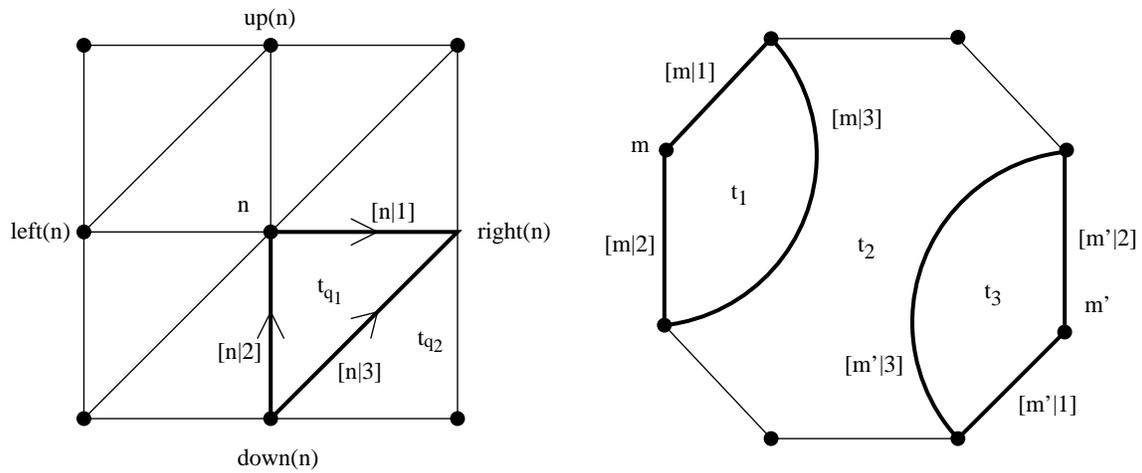,width=15cm}
\vskip 1cm
\caption{The labeling of lattice edges and faces.}
\label{fig5}
\end{figure}

A lattice differential $w$  is determined by its value on the lattice edges. For the lattice $G$ shown in Fig. \ref{fig1} this dependence can be expressed as $w[n|k]: n=1,\ldots,{\cal N};k=1,2,3$ where $[n|1]$ stands for the horizontal edge immediately right of vertex $n$, $[n|2]$ for the vertical edge immediately below and $[n|3]$ for the diagonal edge laying in between, see  Fig. \ref{fig5}. Edges are oriented as shown in  the figure according to their classification into  horizontal, vertical or diagonal. The integral of a lattice differential $w$ along a path  $C$ of lattice edges is defined to be the sum of the values of $w$ on all the edges $e$ included in the path
\be
   \int_{C} w= \sum_{e\in C } \pm \,w[e]
\ee
the minus sign applying  to edges with opposite orientation to the one of the path.

A lattice volume form $\eta$ is determined by its value on each
lattice face. The genus two triangular lattice $G$ has $2{\cal
N}-2$ faces, two of which are hexagons the remaining being
triangles. A volume form $\eta$ in $G$ can be expressed as
$\eta[q]: q=1,\ldots,2{\cal N}-2$ where $q$ is an integer
labelling of the faces of  $G$. The integral of a volume form
$\eta$ over a given area $A$ on the lattice is the sum of the
values that $\eta$ takes on all the lattice faces $q$ contained in
that area \be
   \int\!\!\!\int_{A} \eta= \sum_{q\in A }  \eta[q].
\ee

 The lattice exterior derivative $d$ is a linear operator that takes lattice functions into lattice differentials and lattice differentials into lattice volume forms. It can be defined as
\bs
  (d\, f)\, [e]&=& f[n_{e1}] - f[n_{e2}] \label{der}\\
  (d\, w)\, [q]&=& \sum_{e\in q} \pm w[e] \nn
\es
where $n_{e1}$ and $n_{e2}$  are the two vertices corresponding to the end and beginning of the oriented edge $e$ and the sum runs over all the edges  $e$ in the boundary of the face $q$, the plus or minus sign being determined by the relative orientation of the edge  $e$ to the anticlockwise orientation of the face.

The exterior derivative defined in this way satisfies a discrete version of Stokes theorem. Let $C(n,n')$ be a path of lattice edges from vertex $n$ to vertex $n'$ and $A$ an area on the lattice, then we have that
\bs
   \int_{C(n,n')} d\, f&=& f[n']-f[n]  \\
   \int\!\!\!\int_{A} d\, w&=&  \int_{\partial A} w \nn
\es
where $\partial A$ is the path along the boundary of the area $A$ with an anticlockwise orientation.

An inner product on lattice functions, differentials and forms can be defined as
\bs
    (f,f') &=&  \sum_n   \rho_0(n)\;f[n]\overline{f'[n]} \\ \label{cinn}
    (w,w') &=&  \sum_e   \rho_1(e)\;w[e]\,\overline{w'[e]} \nn \\
    (\eta,\eta ') &=&  \sum_q   \rho_2(q)\;\eta[q]\overline{\eta' [q]} \nn
\es
where the $\rho_i$ are positive functions on the vertices, edges and faces. For our purposes, in the lattice $G$  we define $\rho_0(n)=\rho_2(q)=1$ and let $\rho_1(e)$ take one of three values $\rho_H$, $\rho_V$ and $\rho_D$ for horizontal, vertical or diagonal edges respectively. In terms of the edge labeling introduced above we have then
\be
   (w,w') =  \sum_{n}  \left( \rho_H\;w[n|1]\,\overline{w'[n|1]}+ \rho_V\;w[n|2]\,\overline{w'[n|2]} + \rho_D\;w[n|3]\,\overline{w'[n|3]} \right) \label{inner}
\ee

From a choice of inner product there follows a definition of the lattice co-derivative $\delta $, the  adjoint operator of the exterior derivative,
\bs
   (w,d\,f) &=& (\delta \, w,f) \\
   (\eta,d\,w) &=& (\delta \, \eta,w).\nn
\es
The  co-derivative  takes  lattice differentials into  lattice functions and  lattice volume forms into  lattice differentials and can be expressed as
\bs
   (\delta \,w)\,[n]&=& \sum_{e \ni n} \pm \rho_1(e)\;w[e]\label{del} \\
  (\delta \,\eta\,)\,[e]&=&  \frac{1}{\rho_1(e)}\; (\eta[q_{el}]-\eta[q_{er}]) \nn \\
\es
where the sum on the first equation runs over all edges $e$ that contain the vertex $n$, the plus and minus sign corresponding to $n$ being the ending or the starting vertex of the oriented edge. On the second equation $q_{el}$ and $q_{er}$ stand for the faces left and right of the edge $e$ with respect to its orientation. For the lattice $G$ we have explicitly
\bs
  \delta\,w[n]&=&  \rho_H\,\left( w[\lf(n)|1]-w[n|1] \right)+  \rho_V\,\left( w[n|2]-w[\up(n)|2]\right) \label{coder}\\
              && +  \rho_D\,\left( w[\lf(n)|3]-w[\up(n)|3]\right)  \nn
\es
where  the functions $\ri(n), \lf(n), \up(n)$ and $\dw(n)$  give the label of the vertex immediately right, left, above and below of vertex $n$, see Fig. \ref{fig5}.

Given a definition of the lattice exterior derivative and co-derivative we define, in exact analogy with the continuum concepts, a lattice differential $w$ to be  closed if $d\,w=0$, exact if $w=d\,f$, co-closed if $\delta\,w=0$ and co-exact if $w=\delta\,\eta$.
A lattice differential is said to be harmonic if it is both closed and co-closed
\bs
     w \text{   is harmonic } \equiv \left\{
              \begin{array}{cl}
                d\, w[q] =0 ,   &   \ \text{for all faces  } q \text{  in  } G \\
                \delta \, w[n]=0 ,&    \ \text{for all vertices  } n 
\text{  in  } G.            \end{array}   \right.  \label{harm}
\es

These equations can be seen as linear system of equations on the unknowns $w[n|k]$. We will see that for a genus $g$ lattice there are $2g$ linear independent solutions, corresponding to the $2g$ dimensional vector space of harmonic differentials of the continuum theory.

\subsection{The effective square lattice description}

The linear system (\ref{harm}) can be partially solved by noting that the value of a closed differential on the diagonal edges $w[n|3]$ is determined by its values on the adjacent horizontal  $w[n|1]$ and  vertical $w[n|2]$ edges. Using this constrain to eliminate the $w[n|3]$  from the second equation in (\ref{harm}) the problem of determining the lattice harmonic differentials on the triangular lattice becomes defined on an effective square lattice.

Let $G_\sq$ be the  square lattice obtained from the triangular lattice $G$ by removing all the diagonal edges.
Consider Fig. \ref{fig5}; any square face $q$ of $G_\sq$ contains two  triangular  faces of $G$, $t_{q_1}$ and $t_{q_2}$. These two triangular faces  share the diagonal edge $[n|3]$ that enters the  closeness condition in (\ref{harm})  in two equations
\be
    d\, w[t_{q_1}]=0, \ \ \ \ d\, w[t_{q_2}]=0
\ee
that can be equivalently  expressed as
\bs
    d_\sq\, w[q]&= & 0 \\
      w[n|3]&=&w[n|1]+w[n|2] \label{diag}
\es
where $d_\sq$ is the exterior derivative on the square lattice $G_\sq$ and the second equation determines the value of the differential on the diagonal edge.

In a similar way the hexagonal faces of $G$ can be seen as contained on an octagonal face of $G_\sq$ together with two triangular faces, see Fig. \ref{fig5}, The closeness condition on these three faces of $G$ is
\be
    d\, w[t_1]=0, \ \ \ \ d\, w[t_2]=0, \ \ \ \ d\, w[t_3]=0
\ee
 and can also be expressed in terms of the vanishing of exterior derivative $d_\sq$ on the octagon $q$ and condition (\ref{diag}) for the two diagonal edges involved
\bs
    d_\sq\, w[q]&=&0 \\
    w[n|3]&=&w[n|1]+w[n|2] \ \ \  \text{for } n=m,m'. \nn
\es

Using (\ref{diag}) to eliminate the $w[n|3]$ terms in the co-closeness condition in (\ref{harm}) we obtain that the harmonic differentials on the triangular lattice $G$ are completely determined by the following conditions on their restriction to the squared lattice $G_\sq$
\bs
     w \text{   is harmonic } \equiv \left\{
              \begin{array}{cl}
         d_\sq\, w[q] =0 , &   \ \text{for all faces  } q \text{  in  } G_\sq \\
         \delta_\sq \, w[n]=0 ,&    \ \text{for all vertices  } n \text{  in  } G_\sq
      \end{array}   \right.  \label{harmsq}
\es
where the finite difference  operators on $G_\sq$ are given by
\bs
   d_\sq\, w[q]&=&\sum_i \left( w[\dw(\hat{q}_i)|1] + w[\ri(\hat{q}_i)|2]- w[\hat q_i|1]  - w[\hat{q}_i|2]\right) \label{der0}  \\
  \delta_\sq\,w[n]&=& \rho_H\,\left( w[\lf(n)|1]-w[n|1] \right)+ \rho_V\,\left( w[n|2]-w[\up(n)|2]\right) \label{coder0}\\
              && + \rho_D\,\left( w[\lf(n)|1]+w[\lf(n)|2]-w[\up(n)|1]-w[\up(n)|2]\right)  \nn
\es
the sum in (\ref{der0}) goes over $i=1,2$ for the two octagonal faces of $G_\sq$, with $\hat q_1$ and $\hat q_2$ being the labels of the two vertices that can be seen as the octagon upper left vertices in  Fig. \ref{fig1}; and over $i=1$ for the squared faces of $G_\sq$ of which  $\hat q_1$ is the upper left vertex.

This notation\cite{me} for the faces and vertices of a locally square lattice, has the advantage of  relating  the labeling of vertices with the labeling of faces: if $q$ labels a face then $\hat q_i$  label the upper left vertices of that face. Reciprocally if $n$ labels a lattice vertex we define  $\tilde n$ to be  the label of the face of which $n$ can be seen as the upper left vertex. We then have the relations
\bs
   \tilde{\hat{q}_i}&=&q \\
    n &\in & \{ \hat{\tilde n}_1 ,\hat{\tilde n}_2  \}
\es

\subsection{Evaluation of the harmonic differentials}

For a given lattice $G$ a basis of the space of  harmonic differentials can be evaluated  by  direct solution of one of  the equivalent linear systems (\ref{harm}) or (\ref{harmsq}). 

The number of solutions can be bounded from below by subtracting the number of unknowns by the number of equations. Two additional solutions are provided by the two dimensional space of constant differentials and the total number of solutions is at least
\be
   \text{\#(edges)}-\text{\#(vertices)}-\text{\#(faces)}+2= -\chi+2 =2g
\ee
where $\chi$ is the Euler characteristic of the lattice. There are no additional solutions because harmonic differentials are completely determined by its periods, or integrals, along the non-trivial homology cycles of the lattice and $2g$ is precisely the dimension of the first homology group (see\cite{me} for further details).

The linear system (\ref{harmsq})  can always be evaluated numerically, although the evaluation of the kernel of a large linear system is computationally demanding. For practical purposes there is a much more efficient method based on a discrete version of the Hodge decomposition theorem.
This theorem states that the  space of lattice differentials  in $G_\sq$ has an orthogonal decomposition in terms of exact, co-exact and harmonic differentials and that any  lattice differential $w$ can be written in an unique way as
\be
 w= d_\sq f+ h+\delta_\sq \eta \label{hdg}
\ee
where $h$ is a harmonic. The orthogonality of the different kinds of differentials follows directly from (\ref{harmsq}) and the property $d_\sq d_\sq=\delta_\sq \delta_\sq=0$.

\begin{figure}[h]
\vskip 3cm
\epsfig{figure=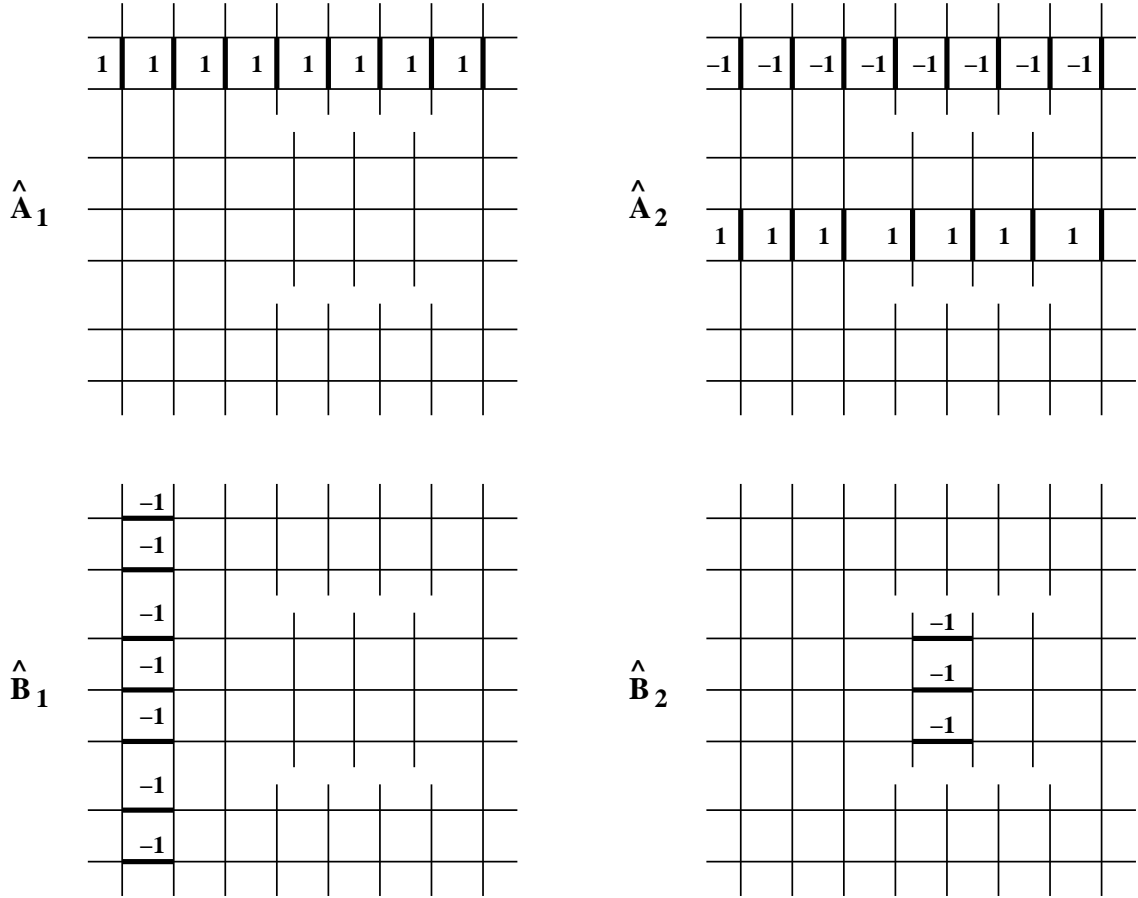,width=15cm}
\vskip 1cm
\caption{The closed differentials $\hat A_k,\hat B_k$. The differentials are zero on all edges except the ones crossed by the respective $\tilde b_k$, $\tilde a_k$ loop. These edges are shown in bold on the figure, together with the value that the differential takes at that edge. For convenience we use a $\tilde b_2$ loop different but homologically equivalent to the one shown in Fig. \ref{fig1}.}
\label{fig6}
\end{figure}

This result allows the determination of harmonic differentials by orthogonal projections. Our objective is to obtain a basis $\{A_k, B_k :\;  k=1,2\}$ of the space of harmonic differentials satisfying the normalization conditions
\bs
    \int_{a_j} A_k&=& \delta_{kj},\ \ \int_{b_j} A_k= 0 \label{harmbasis}\\
    \int_{a_j} B_k&=& 0,\ \ \; \int_{b_j} B_k=  \delta_{kj} \nn
\es
with the $a_j, b_j$ being any choice of closed paths on the lattice representing a basis of the first homology group.
 We proceed in the following way: start with closed but not harmonic differentials $\hat{A}_k,\hat{B}_k$ with the periods required by (\ref{harmbasis}). A possible choice for these differentials is shown in Fig. \ref{fig6} where we take $\hat{A}_k$ ($\hat{B}_k$) to be zero on all edges except the ones crossed by the dual lattice cycles $\tilde{b}_k$ (respectively $\tilde{a}_k$).
Since closed differentials are orthogonal to co-exact differentials it follows from (\ref{hdg}) that closed differentials can be written as the sum of a harmonic differential with the same periods and an exact differential
\bs
      \hat{A}_k&=&A_k+d_\sq f^{A}_k \label{decomp}\\
      \hat{B}_k&=&B_k+d_\sq f^{B}_k \nn
\es
 applying  $ \delta$ on the these  equations and using (\ref{harmsq})  we obtain
\bs
  \Delta_\sq\, f^{A}_k &=& \delta_\sq\,\hat{A}_k \label{eqnum} \\
  \Delta_\sq\, f^{B}_k &=& \delta_\sq\,\hat{B}_k  \nn
\es
where the Laplacian operator $\Delta_\sq= \delta_\sq d_\sq$ on lattice functions is given explicitly by
\bs
   (\Delta_\sq\;f)[n]&=& 2(\rho_H +\rho_V +\rho_D )\,f[n]\label{lap}\\
                 && -\rho_H\,(f[\ri(n)]+f[\lf(n)]) \nn \\
                 && -\rho_V\,(f[\up(n)]+f[\dw(n)]) \nn \\
                 && -\rho_D\,(f[\ri(\up(n))]+f[\dw(\lf(n))]). \nn
\es

The Laplacian operator can be seen as a ${\cal N}\times {\cal N}$ matrix of rank $({\cal N}-1)$ acting on the functions ${\cal N}$-vector. The linear system of equations (\ref{eqnum}) can then be solved numerically for reasonably large ${\cal N}$ by fixing the value of the functions  at a lattice vertex and the  solutions can be  differentiated and subtracted from the $\hat{A}_k,\hat{B}_k$ to obtain a normalized basis of the space of harmonic differentials $\{A_k, B_k :\;  k=1,2\}$. This method is numerically more efficient than the direct solution of the linear system (\ref{harmsq}).

\section{Holomorphic Differentials and the Period Matrix}\label{section4}

The second step in the definition of a lattice period matrix is the decomposition of the space of lattice harmonic differentials, obtained in the previous section, into  holomorphic and  anti-holomorphic sub-spaces.
For general values of the edge weights $\rho_H$, $\rho_V$ and $\rho_D$ in (\ref{inner}) this is not possible. However if the edge weights are related through the Ising model criticality condition we are able to interpret  $d_\sq$ and $\delta_\sq$  as the geometrical realization of a square lattice tilted by an angle $\theta$ and a generalization of the square lattice formalism\cite{me} allows the definition of a discrete period matrix for the triangular lattice $G$.

\subsection{Geometrical interpretation of the criticality condition}

In order to reproduce the period matrix in the Ising  model finite size correction (\ref{shape}) the edge weights  $\rho_H$, $\rho_V$ and $\rho_D$  must be given in terms of the Ising model coupling constants along the same direction as
\be
     \rho_j=\sinh{2 K_j} \label{param}.
\ee
Since the coupling constants at  criticality verify (\ref{critic}) we can parameterize the edge  weights $\rho_i$ in terms of a positive number  $r$ and an angle $\pi/2 <\theta \leq \pi$
\bs
    \rho_H&=& \frac{1}{\sin{\theta}} \left( r + \cos{\theta} \right) \label{para} \\
    \rho_V&=& \frac{1}{\sin{\theta}} \left( \frac{1}{r} + \cos{\theta} \right) \nn\\
    \rho_D&=& \frac{- \cos{\theta}}{\sin{\theta}} \nn
\es
with
\bs
   -\cos{\theta}&=& \frac{\rho_D}{\sqrt{(\rho_H+\rho_D)(\rho_V+\rho_D) }} \label{coshv}\\
    r&=& \sqrt{\frac{(\rho_H+\rho_D)}{(\rho_V+\rho_D)}}.
\es

\begin{figure}[h]
\vskip 3cm
\epsfig{figure=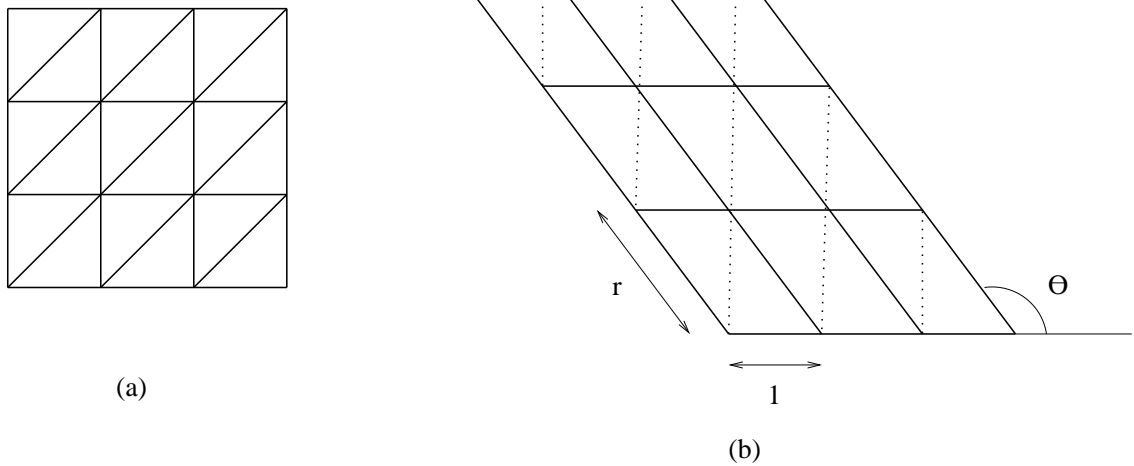,width=15cm}
\vskip 1cm
\caption{The aspect ratio $r$ and tilt angle $\theta$. a) the combinatorial description of the lattice. b) the geometrical description of lattice from the parametrization of $\sinh{2K_i}$.}
\label{fig70}
\end{figure}

This parameterization has a clear geometrical interpretation with
$r$ being the ratio of length scales along the vertical and
horizontal directions and $\theta$ a tilt angle, see Fig.
\ref{fig70}. To show this, consider the co-derivative
(\ref{coder0}) in terms of the new parameters,  dropping an
overall $\sin{\theta}$ factor, we have that \bs
   \delta_\sq\;f[n]&=& r\ \left( w[\lf(n)|1]-w[n|1]\right) + 1/r\ \left(w[n|2]-w[\up(n)|2]\right) \label{coder02} \\
                 && -\cos{\theta}\ \left( w[n|1] - w[\up(n)|1] + w[\lf(n)|2] - w[n|2] \right ). \nn
\es
and the discrete Laplacian operator on lattice functions $(\Delta_\sq=\delta_\sq d_\sq )$   is given by
\bs
   (\Delta_\sq\;f)[n]&=& 2(r+1/r+\cos{\theta})\,f[n]\label{lap2}\\
                 && -(r+\cos{\theta})\,(f[\ri(n)]+f[\lf(n)]) \nn \\
                 && -(1/r+\cos{\theta})\,(f[\up(n)]+f[\dw(n)]) \nn \\
                 && +\cos{\theta}\,(f[\ri(\up(n))]+f[\dw(\lf(n))]) \nn
\es

Modulo a constant term, this Laplacian is the discretization of a continuum Laplacian describing a geometry with an angle $\theta$ between the directions $x_0$ and $x_1$
\be
   \Delta=-\frac{1}{\sin^2(\theta)}\left( \frac{\partial^2}{\partial x_0^2} -2\cos{\theta}\frac{\partial^2}{\partial x_0\partial x_1}+ \frac{\partial^2}{\partial x_1^2}  \right).
\ee
with $r$ being the ratio of the lattice spacings along these directions, precisely the geometry of Fig. \ref{fig70}.

The Ising model criticality condition (\ref{critic}) allows the description of discrete harmonic differentials on a triangular lattice as harmonic differentials on a square lattice tilted by an angle $\theta$. While the Ising criticality condition has been previously given  a geometrical interpretation\cite{mercat} at the level of discrete holomorphy, the reasoning above shows that a geometrical interpretation exists already at the level of discrete harmonicity.

\subsection{The Hodge operator}

In the continuum theory  the holomorphic and anti-holomorphic sub-spaces are obtained from the space of harmonic differentials by the projection operators
\be
P=(1 + i\bigstar)/2 \ \ \text{and} \ \ \bar P=(1 - i\bigstar)/2 \label{proj}
\ee
where  $\bigstar$ is the (continuum) Hodge operator\cite{farkra,hodge}. This operator is an endomorphism on the space of  harmonic differentials and satisfies \mbox{$\bigstar^2=-1$}. Its action on differentials is given by
\be
     \bigstar\left[f_x(x,y)\, dx + f_y(x,y) dy\right]= -f_y(x,y)\, dx + f_x(x,y) dy
\ee
and a differential is said to be holomorphic if it is of the form $P w$ with $w$ harmonic.

We will proceed in a similar way in the discrete theory and define lattice holomorphic differentials  by a projection with a discrete Hodge operator $\star$. This  operator on lattice differentials should be defined in such way that $(1\pm i\star)/2$ are projection operators and endomorphisms on the space of harmonic differentials, or equivalently that
\bs
    \left\{ \begin{array}{rl}
                    d\, w[q] =0 , &  \text{ for all  } q\\
                \delta\, w[n]=0 ,   & \text{ for all  } n\end{array}   \right.
    &\ \ \Rightarrow\ \ &
    \left\{ \begin{array}{cl}
                    d\,\star\, w[q] =0 , &\text{ for all  } q\\
                \delta\,\star\, w[n]=0,   & \text{ for all  } n \end{array}   \right.\label{cond1} \\
   \star\star&=&-1 \label{cond2}
\es

There is no known definition of a discrete Hodge operator that satisfies these properties precisely while keeping the dimensions of the space of holomorphic differentials equal to $g$. We propose the following definition

\textbf{Definition:} the Hodge operator on the effective square
lattice  $G_\sq$ with edge weights parameterized according to
(\ref{para}) is defined by

\bs
  (\star w)\,[n|1] &=& -\frac{1}{r}\frac{1}{\sin{\theta}}\, w[\up(n)|2]+\frac{\cos{\theta}}{\sin{\theta}}\,  w[\up(n)|1]  \label{star}\\
  (\star w)\,[n|2] &=& \ \ r\frac{1}{\sin{\theta}}\, w[\lf(n)|1]-\frac{\cos{\theta}}{\sin{\theta}}\,  w[\lf(n)|2] \nn
\es

For $\theta=\pi/2$ we recover the  discrete Hodge star for the simple square lattice\cite{me}, minor differences are due to the different labeling we have chosen for the lattice edges, see Fig. 9 in that reference.

This definition of the discrete Hodge operator relates the exterior derivative $d$ with the co-derivative $\delta$ in a way that  closely satisfies (\ref{cond1}). From the definition above, (\ref{der0}) and (\ref{coder02}) it follows that
\bs
  (d\star\! w)[q]&=& - \frac{1}{\sin{\theta}} \sum_i \delta \, w[\hat{q}_i]  \label{ddstar1} \\
  (\delta \star\! w)[n]&=& \left\{
              \begin{array}{l}
        \alpha(n)   \text{   if }  n \in \{n_1,n_2,n_3,n_4\}\\
        \sin{\theta}\,d\, w[\widetilde{\up}(\lf(n))] + \frac{\cos{\theta}}{\sin{\theta}}(\delta\, w[\up(n)]- \delta\, w[\lf(n)]) \text{   otherwise }
              \end{array}   \right. \label{ddstar2}
\es
where in the first equation the sum goes over $i$=1,2 for the octagonal faces and over $i$=1 for squared faces. In the second equation, the $n_k$ are the four lattice vertices where ${\lf}(\up(n_k))\neq{\up}(\lf(n_k))$ and $\widetilde{\up}(n)$ stands for the tilde of the vertex $\up(n)$.

Because of the four vertices $n_k$ the discrete Hodge operator
fails to be an endomorphism on the harmonic lattice differentials:
the requirement that a differential $w$ is closed and co-closed is
not sufficient to ensure that the corresponding  $\star w$ is
co-closed. This is due to the fact that the lattice has more
vertices than faces and co-closeness is a condition verified on
vertices. The condition is still partially satisfied in the sense that, if
$n_1$ and $n_2$ are on the same octagon,  then the $\alpha(n_k)$ satisfy 
\bs
    \alpha(n_1)+\alpha(n_2)&=& \sin{\theta}\,d\, w[\widetilde{\up}(\lf(n_1))] \\
     &&+ \frac{\cos{\theta}}{\sin{\theta}}(\delta\, w[\up(n_1)]- \delta\, w[\lf(n_1)]+\delta\, w[\up(n_2)]- \delta\, w[\lf(n_2)])  \nn
\es
and similarly for  $\alpha(n_3)+\alpha(n_4)$.

Unlike the continuum Hodge star, our lattice definition does not satisfy (\ref{cond2}) but rather
\bs
  \star^2 w[n|1]&=& -w[\lf(\up(n))|1] + \ \frac{\cos{\theta}}{\sin{\theta}}\,\left( D(\star w)[n]+ \  \frac{\cos{\theta}}{\sin{\theta}}\, D(w)[\lf(n)]\right) \label{sqstar}  \\
  \star^2 w[n|2]&=& -w[\up(\lf(n))|2] +\  r\frac{{\cos{\theta}}}{\sin^2{\theta}}\, D(w)[\lf(n)]\nn
\es
were $D(w)$ is defined as
\be
   D(w)[n]= (w[\up(n)|1]-w[\lf(n)|1]) - \frac{1}{r}\cos{\theta}\,(w[\up(n)|2]-w[\lf(n)|2]).
\ee 
as in the simple square lattice case\cite{me} the action of  $\star^2$ on lattice differentials produces a diagonal translation, the size of a lattice spacing, that becomes negligible in the continuum limit. 

\subsection{ Differentials in the dual lattice}

We can also consider the construction of harmonic and holomorphic
differentials on $G^*_\sq$, the dual lattice of $G_\sq$. While most
of the concepts above follow through with minor changes there are
a few relevant differences that make the dual lattice a testing
ground for the dependence of our definitions on the lattice
structure. 

We define  $G^*_\sq$ as the dual of $G_\sq$ in the
usual graph theoretical sense together with the same assignment of
edge weights, parameterized by $r$ and $\theta$, for horizontal
and vertical edges.
The two lattices $G_\sq$ and  $G^*_\sq$ differ at the conical singularities: where one has an octagonal face the other has a vertex with eight  adjacent edges. This fact has important consequences in equations  (\ref{ddstar1}) and (\ref{ddstar2}). As mentioned before closed differentials $w$ in  $G_\sq$ do not generate  co-closed differentials of the form $\star w$  because the lattice has more vertices than faces.  For the dual lattice $G^*_\sq$, where the number of faces is larger than the number of vertices, the opposite happens and (\ref{ddstar1})  is given in terms of $\alpha(n)$ terms not expressible in terms of derivatives and co-derivatives at a single point.

\begin{figure}[h]
\vskip 3cm
\epsfig{figure=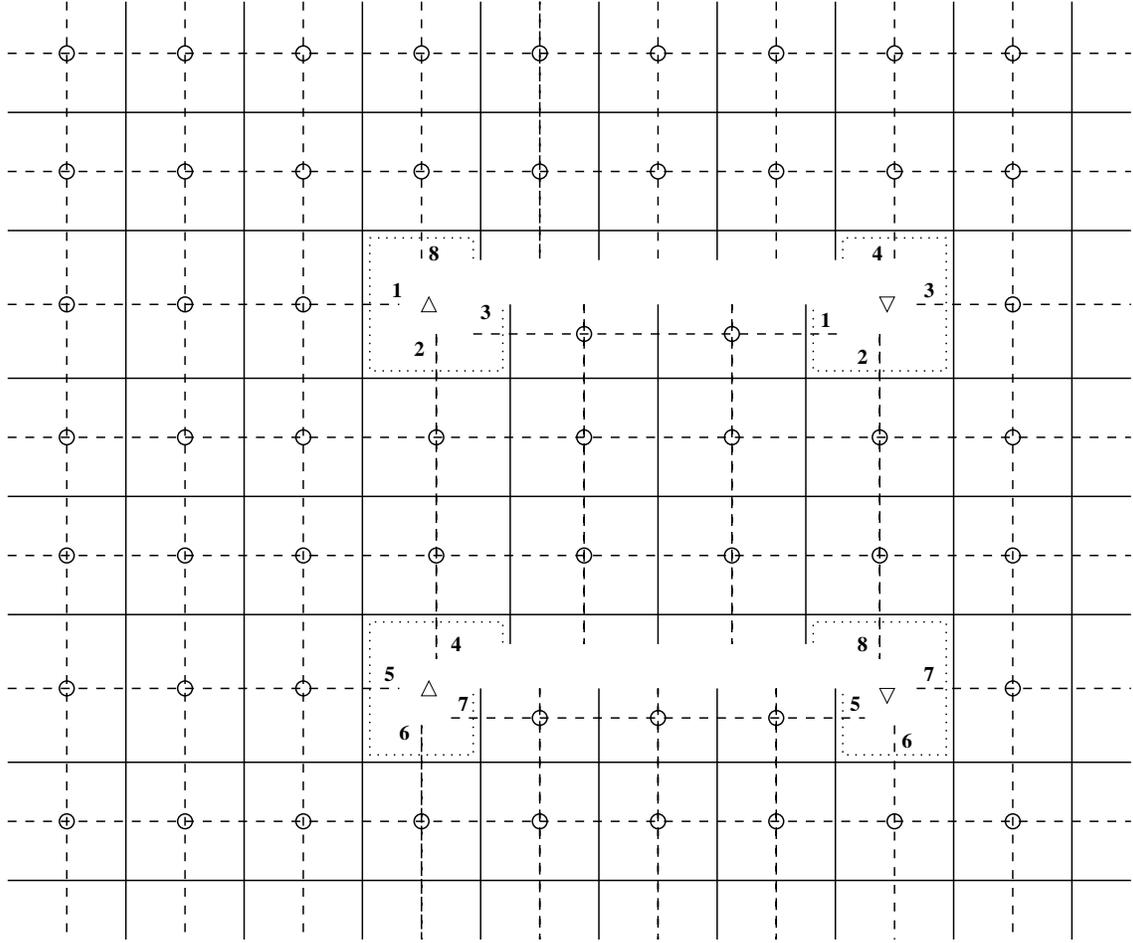,width=15cm}
\vskip 1cm
\caption{The dual lattice $G^*_\sq$ is shown, in dashed lines, together with the direct lattice $G_\sq$, on full lines. Each circle is a dual lattice vertex. The two up triangles correspond to a single dual lattice vertex, equally for the two down triangles. The numbers labeling the edges that enter the octagonal faces show the order in which they join the corresponding vertex.}
\label{fig7a}
\end{figure}

\begin{figure}[h]
\vskip 3cm
\epsfig{figure=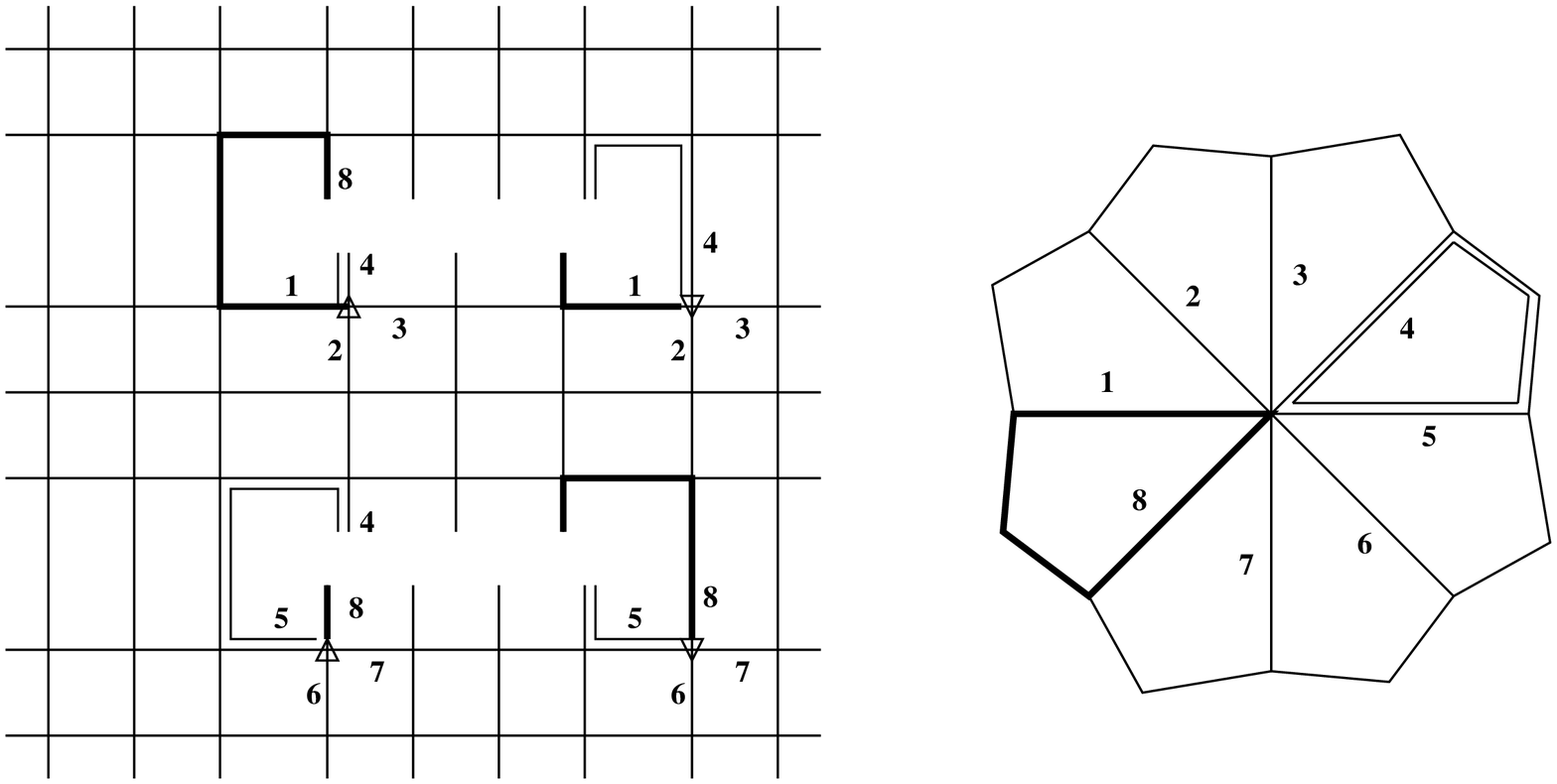,width=15cm}
\vskip 1cm
\caption{The dual lattice $G^*_\sq$ schematically represented as $G_\sq$ with two vertices identified on each octagon. On the right the local arrangement of the dual edges on the octagonal faces is shown.}
\label{fig7b}
\end{figure}

For calculation purposes the dual lattice $G^*_\sq$ can be schematically represented  as $G_\sq$ with two vertices identified on each octagonal face , see Figs. \ref{fig7a} and \ref{fig7b}. The faces of $G^*_\sq$ are all squares and two of the lattice vertices have eight adjacent edges. Since the edges of $G^*_\sq$ are in one to one correspondence with the edges of $G_\sq$ and we can use a edge labeling for $G^*_\sq$ similar to the one used for $G_\sq$.
The Hodge operator and the exterior derivative on functions can be defined in the same way for both lattices while the exterior derivative $d^*_\sq$ and the co-derivative $\delta^*_\sq$  are defined as

\bs
   d^*_\sq\, w[q^*]&=&\left( w[\hat q^*|1] + w[\ri(\hat q^*)|2] - w[\up(\hat q^*)|1] - w[\hat q^*|2]\right) \label{derd}  \\
  \delta ^*_\sq\,w[n^*]&=& \sum_i \rho_H\,\left( w[\dw(n^*_i)|1]-w[n^*_i|1] \right)+ \rho_V\,\left( w[n^*_i|2]-w[\up(n^*_i)|2]\right) \label{coderd}\\
              && + \rho_D\,\left( w[\lf(n^*_i)|1]+w[\lf(n^*_i)|2]-w[n^*_i|1]-w[n^*_i|2]\right)  \nn
\es
where in the first equation $\hat q^*$ is the vertex of  $G^*_\sq$ that is the upper left corner of the square face $q^*$ of $G^*_\sq$.
 In the second equation the sum goes over  $i=1,2$ when $n^*$ is one of the $G^*_\sq$ vertices with eight edges and $i=1$ for all other vertices.
The functions $\ri(n), \lf(n), \up(n)$ and $\dw(n)$  and the relations between vertices, edges and faces in  $G^*_\sq$ are evaluated on  $G_\sq$ with values in  $G^*_\sq$ which means that we start by distinguishing over different $n^*_i$ but identify them in the final result. This set of rules provides a complete prescription to evaluate holomorphic differentials in  $G^*_\sq$.

\subsection{The torus modular parameter}

As an illustration of the formalism developed above we will consider the genus one toroidal lattice and reproduce the  modular parameter, known by exact asymptotic analysis\cite{nashoco2}.

 Consider a toroidal triangular lattice with $M$ rows and $N$ columns, characterized by the coupling constants $K_h$, $K_v$ and $K_d$ along the horizontal, vertical and diagonal edges. Let $\{a,b\}$ be a basis of the first homology group with $a$ being a horizontal loop and $b$ a  vertical loop. The harmonic differentials, solution of (\ref{harmsq}), on this lattice are the constant differentials
\be
 \begin{array}{ll}
   A[n][1]=\alpha, &  A[n][2]=\beta
 \end{array}
\ee
for all vertices $n$, with  $\alpha$ and $\beta$ being two arbitrary constants. These harmonic differentials can be projected into the space of holomorphic differentials as
\bs
   (1+i \star)\, A[n][1]&=& \alpha +i \left( \frac{\cos{\theta}}{\sin{\theta}}\, \alpha -\frac{1}{r}\, \frac{1}{\sin{\theta}}\, \beta\right) \\
   (1+i \star)\, A[n][2]&=& \beta +i \left( - \frac{\cos{\theta}}{\sin{\theta}}\, \beta +r\, \frac{1}{\sin{\theta}}\, \alpha   \right)
\es
for all vertices $n$.
The lattice modular parameter $\tau$ is obtained by integrating this differential along a loop of the homology class of $b$ after a proper normalization is chosen
\bs
  \int_{a} (1+i\star) A &=& 1 \\
  \int_{b} (1+i\star) A &=&\tau
\es
immediately we reproduce the know result\cite{nashoco2}
\be
   \tau= \frac{M}{N}\,r\,\left(\cos{\theta}+i\,\sin{\theta}\right)
\ee
with $r$ and $\theta$ given by (\ref{param}) and (\ref{coshv}).

\subsection{The  period matrix on higher genus lattices}

After obtaining a basis  $\{A_k, B_k\}$ for the space of harmonic differentials on a given lattice $G$, using the procedures of section \ref{section3}, we proceed to evaluate a basis for the space of holomorphic differentials of the form \mbox{$\{\Gamma_k=(1+i\star)H_k : k=1\ldots,g\}$} where the $H_k$ are harmonic differentials, satisfying the normalization conditions
\bs
    \int_{a_k} (H_l+i\star H_l) &=& \delta_{kl} \label{holo} \\
     \int_{b_k}  (H_l+i\star H_l) &=& \Omega_{kl} \label{period}.
\es

Equation (\ref{holo}) is a set of $2g$ real constraints on $H_k$ from which we can determine the coefficients of  $H_k$ in the basis  $\{A_k, B_k\}$ and equation (\ref{period}) is our definition of the finite size lattice period matrix.

The same construction can be made on the dual lattice $G^*_\sq$.
We start by obtaining  a basis of the harmonic differentials
$\{A^*_k, B^*_k :\;  k=1,2\}$   using the procedure of section
\ref{section3} and (\ref{derd}) and  (\ref{coderd}). Then using
the same projection operator we obtain a basis for the space of
differentials \mbox{$\{\Gamma^*_k=(1+i\star)H^*_k :
k=1\ldots,g\}$} with the $H^*_k$ being harmonic differentials in
$G^*_\sq$, satisfying the normalization conditions \bs
    \int_{a_k} (H^*_l+i\star H^*_l) &=& \delta_{kl} \label{holo1} \\
     \int_{b_k}  (H^*_l+i\star H^*_l) &=& \Omega^*_{kl} \label{period1}
\es
The  discrete period matrix on the dual lattice $\Omega^*$ is not in general equal to that of the direct lattice $\Omega$ as we will see in the next section by explicit computation.

\subsection{Connection with other formulations}

The definitions we introduced in this section treat the direct
lattice and the dual lattice separately, using a discrete Hodge
operator defined exclusively on either the direct lattice or on
the dual lattice. This was possible due to the reduction of the
problem to a tilted square lattice, that can be seen as
approximately self-dual.

In the literature\cite{books,kotiuga,bourdlala,tarketboss,teix1,hiptmair}
many examples can be found in which the Hodge operator on
d-dimensional lattices is defined as an operator taking p-cochains
on the lattice to (d-p)-cochains on the dual lattice.  For two
dimensional problems, where the Hodge operator is closely related with
the concept of holomorphy, such operators have been extensively
studied by Mercat\cite{mercat,mercat3}, who defines  the Hodge
operator in terms of the double lattice, the direct sum of the
direct lattice $G$ and the dual lattice $G^*$.

In the double lattice formulation the number of holomorphic differentials is double of the continuum analogue and we lose either the interpretation of the period matrix as the periods of a holomorphic basis or its expected dimension. A $2g\times 2g$ period matrix can  be defined\cite{mercat3} using a combination of (\ref{holo}) and (\ref{period}) together with a procedure that yields a $g\times g$ period matrix.

For a square, $\rho_D=0$, version of the lattices $G$ that we study in this paper the $\Omega$ and $\Omega^*$ period matrices we introduced are closely related with two sub-matrices of the $2g\times 2g$ period matrix of the double lattice formulation.
The same is not true for general triangular lattices $G$. Since we require the number of holomorphic differentials to be $g$, and use an exact mapping for discrete harmonic differentials from $G$ to $G_\sq$, our discussion refers to the dual of the effective square lattice $G_\sq$ and not to the dual of the original lattice $G$.

\section{Numerical evaluation of the period matrix}\label{section5}

The formalism developed on the previous sections can be illustrated by the numerical evaluation from first principles of the period matrices in (\ref{resultati}).

\begin{figure}[h]
\vskip 3cm
\epsfig{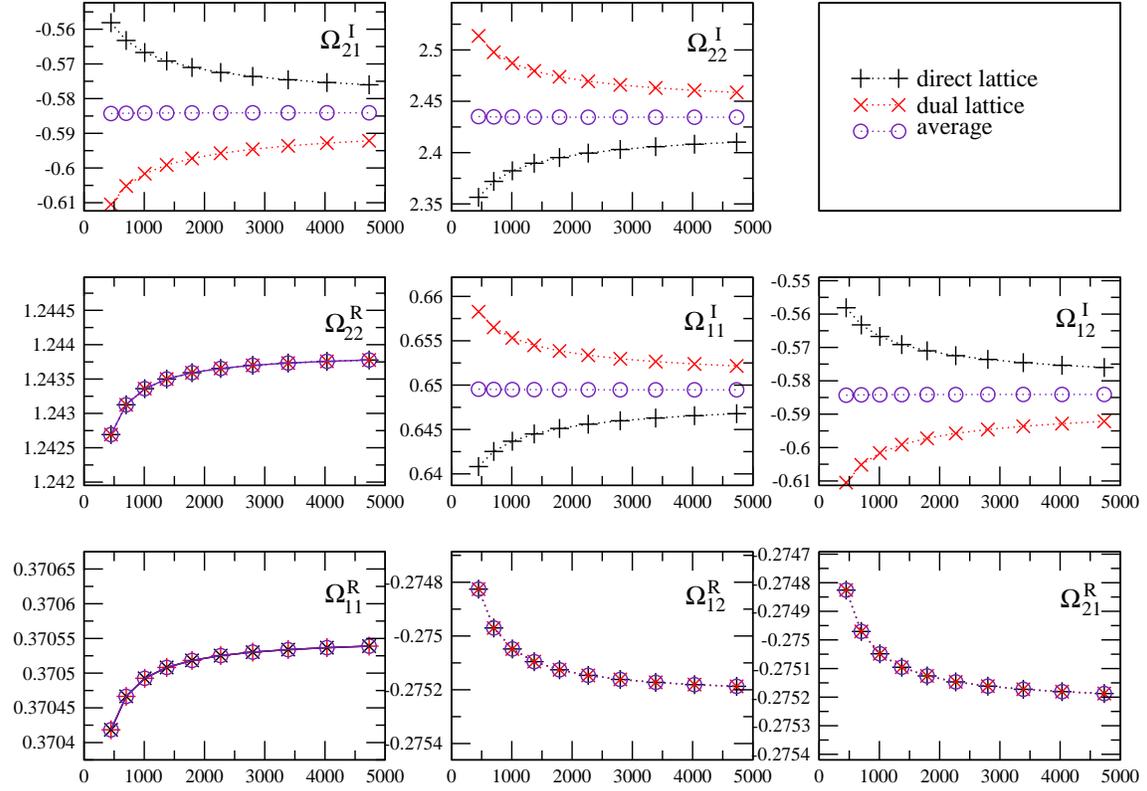}
\vskip 1cm
\caption{The period matrices   $\Omega$ and $\Omega^*$ and their average  as a function of the number of lattice vertices $\cal N$, for the lattice B considered on Figs. \ref{fig3} and \ref{fig4}.}
\label{fig8}
\end{figure}

We consider lattices  with sizes  $(M_1,M_2,K,N_1,N_2)= L(m_1,m_2,k,n_1,n_2)$  for fixed  aspect ratios $(m_1,m_2,k,n_1,n_2)$ and increasing $L$.
The numerically evaluated period matrices for both the direct and the dual lattice, $\Omega$ and  $\Omega^*$, were found to be symmetric with positive definite imaginary part for all lattice sizes and coupling constants.
A typical example of the dependence of the entries  of $\Omega$ and $\Omega^*$ on the lattice size  ${\cal N}$ is shown in Fig. \ref{fig8}.

The two period matrices were found to have equal real parts for all couplings and aspect ratios tested while the imaginary part is typically different but converges towards the same value in the large ${\cal N}$ limit.
In the example of Fig. \ref{fig8} we see that the imaginary parts of  $\Omega$ and $\Omega^*$  converge slower  to the continuum limit than their real part, while the average of the imaginary parts converges as fast as the real part.

The period matrix characterizing a given lattice is not uniquely defined but depends on the choice of the basis of the first homology group, $\{a_i,b_i\}$ used for its evaluation. Period matrices evaluated with different bases are related by modular transformations. Consider the  basis $\{a_i,b_i\}$ shown in Fig. \ref{fig1}, we will call this basis H1 and H2 to the basis obtained from this one by exchanging the labeling of the $a_i$ and $b_i$ cycles.

In table \ref{table2} we give examples of  $\Omega$ and  $\Omega^*$,  evaluated  at finite  ${\cal N}$  and the corresponding ${\cal N}\rightarrow \infty$ extrapolation, using the two basis H1 and H2.
In table  \ref{table3} we plot the ratios of theta functions $R_i$ for these period matrices, they are related by a permutation of the index labeling the half-integer characteristic.
From this table we see that modular invariance is broken at finite size by $\Omega$ and $\Omega^*$ but it is satisfied by the average  $(\Omega+\Omega^*)/2$ with a  precision of $1\%$ or better.
In the large  ${\cal N}$ limit both $\Omega$ and $\Omega^*$ satisfy modular invariance approximately while the  average period matrix satisfies it with a precision of $10^{-6}$.

 Similar averages have been considered in the literature\cite{mercat3} but the fact that the average of the two period matrices is modular invariant at finite size and provides a faster convergence to the continuum limit is an unpredicted numerical finding.

The comparison between the averaged period matrix and the period matrices obtained by fitting the ratios of determinants  in (\ref{resultati}), is done in table \ref{table1}. There is a remarkable agreement between the two sets of values. The theta function ratios  corresponding to the two period matrices agree with a precision of $10^{-3}$ or better.
A typical example of the convergence of ratios of determinants and ratios of theta functions for  $\Omega$ and $\Omega^*$ is shown in Fig. \ref{fig3}.

\section{Conclusions} \label{section6}

In this paper we studied a shape and topology dependent finite size correction to the Ising model partition function on genus two triangular lattices.

We found that, in a similar way to the toroidal case\cite{ferdfish,nashoco2}, the determinants of the Kasteleyn adjacency matrices converge at criticality  to a common bulk  term times theta functions of half-integer characteristic.
This result  (\ref{resultati}) together with the Kasteleyn expression for the higher genus partition function (\ref{zkast}) suggests that the shape dependent correction (\ref{shape}) is universal for the Ising model on orientable two dimensional lattices. Different topologies correspond to sums over different genus theta functions while different lattice aspect ratios and coupling constants change only the period matrix $\Omega$.  The analysis done in this paper for genus two  can be readily generalized for arbitrary genus since both the Kasteleyn formalism and the period matrix evaluation are defined independently of the lattice genus.

The period matrix characterizing the lattice continuum limit can be extracted from the finite size correction (\ref{shape}). This provides a unique testing ground for discrete formulations of holomorphy and Riemann surface theory.
We used such a formulation to reproduce the  dependence of the period matrix on the lattice aspect ratios and coupling constants. The present discussion and the recent interest\cite{mercat,mercat3,kenyon} in discrete holomorphy may provide a starting point for a deeper mathematical understanding of the relation between discrete models and certain aspect of conformal field theory.

We also emphasize that the observed dependence of the determinants of
adjacency matrices on the Kasteleyn orientations is precisely the
dependence of the determinant of the Dirac operator on the Riemann
surface spin structures\cite{alvarez1,verver1} of the
corresponding conformal field theory. Such  a one to one
correspondence between discrete lattice determinants and
functional determinants 
is reminiscent of the  Ray-Singer theorem\cite{raysin,muller}
relating   determinants of the analytical Laplacians on a surface
with the determinants of combinatorial Laplacians on a
triangulation of that surface. We believe that further progress
can be made along these directions.

\bigskip
\centerline{{\bf Acknowledgments}}
\bigskip

 This work was partially supported by the Funda\c{c}\~ao para a Ci\^encia e Tecnologia (Portugal) under Grant BD 11503 97, by the National Science Foundation (USA) under Grant No. DMR-0073058, by the EU grant HPRN-CT-1999-00161 and by the ESF network no. 82. Both authors profited from useful discussions with C. Mercat.

\centering
\tightenlines

\pagebreak

\pagebreak

\pagebreak

\pagebreak

\pagebreak

\pagebreak

\pagebreak

\pagebreak

\pagebreak

\pagebreak


\begin{table} \centering
\begin{tabular}{|c|ccc|ccc|ccc|}
 & \multicolumn{3}{c|}{\textbf{A}: [0.426,0.291] (2,2,2,2,2)}   & \multicolumn{3}{c|}{\textbf{B}: [0.369,0.159](4,2,2,2,4)}   & \multicolumn{3}{c|}{\textbf{C}: [0.082,0.291](4,2,2,4,2)}  \\ \hline
i &   $R^2_i$ &$\theta^4_i$(fit)&$\theta^4_i$(eval)  &   $R^2_i$ &$\theta^4_i$(fit)&$\theta^4_i$(eval) &   $R^2_i$ &$\theta^4_i$(fit)&$\theta^4_i$(eval)\\ \hline
 1& 0.15448     & 0.15445  & 0.15368  & 0.01344     & 0.01344  & 0.01341  & 0.14675     & 0.14677  & 0.14512 \\
 2&  2.7$10^{-6}$&    0     &    0     & 1.5$10^{-6}$&    0     &    0     & 8.6$10^{-7}$&    0     &    0    \\
 3&  2.2$10^{-6}$&    0     &    0     & 2.0$10^{-7}$&    0     &    0     & 4.4$10^{-6}$&    0     &    0    \\
 4&  2.2$10^{-6}$&    0     &    0     & 1.6$10^{-6}$&    0     &    0     & 1.0$10^{-6}$&    0     &    0    \\
 5&  5.7$10^{-7}$&    0     &    0     & 6.7$10^{-8}$&    0     &    0     & 2.4$10^{-7}$&    0     &    0    \\
 6&  0.87017     & 0.87020  & 0.87184  & 0.35920     & 0.35921  & 0.36085  & 0.89475     & 0.89477  & 0.89528 \\
 7&  0.00048     & 0.00048  & 0.00045  & 0.00005     & 0.00005  & 0.00005  & 0.00117     & 0.00110  & 0.00111 \\
 8&  0.76230     & 0.76227  & 0.76487  & 0.37222     & 0.37222  & 0.37388  & 1.00000     & 1.00000  & 1.00000 \\
 9&  2.7$10^{-6}$&    0     &    0     & 5.2$10^{-7}$&    0     &    0     & 3.2$10^{-6}$&    0     &    0    \\
10&  0.15448     & 0.15445  & 0.15367  & 1.00000     & 1.00000  & 1.00000  & 0.36001     & 0.36003  & 0.36012 \\
11&  0.18531     & 0.18534  & 0.18381  & 0.03254     & 0.03254  & 0.03222  & 0.10945     & 0.10944  & 0.10852 \\
12&  0.18531     & 0.18534  & 0.18380  & 0.96246     & 0.96245  & 0.96271  & 0.32775     & 0.32773  & 0.32835 \\
13&  5.7$10^{-7}$&    0     &    0     & 6.2$10^{-7}$&    0     &    0     & 1.5$10^{-6}$&    0     &    0    \\
14&  0.87017     & 0.87020  & 0.87183  & 0.81775     & 0.81775  & 0.81676  & 0.69481     & 0.69483  & 0.69485 \\
15&  0.03959     & 0.03956  & 0.03877  & 0.03799     & 0.03799  & 0.03764  & 0.04184     & 0.04181  & 0.04119 \\
16&  1.00000     & 1.00000  & 1.00000  & 0.78572     & 0.78572  & 0.78496  & 0.62338     & 0.62336  & 0.62345 \\\hline\hline
 &$\Omega_{11}$&$\Omega_{12}$&$\Omega_{22}$&$\Omega_{11}$&$\Omega_{12}$&$\Omega_{22}$&$\Omega_{11}$&$\Omega_{12}$&$\Omega_{22}$ \\\hline
 $\Omega^r$(fit)   &  0.2536 & $-$0.1837 & 0.3675 & 0.3697 & $-$0.2754 & 1.2506 &  1.1846 & $-$0.9445 & 1.9637  \\
 $\Omega^i$(fit)   &  1.4042 & $-$1.1978 & 2.3956 & 0.6491 & $-$0.5826 & 2.4293 & 1.1483 & $-$1.0984 & 2.5132  \\  \hline
 $\Omega^r$(eval)   &  0.2549 & $-$0.1855 & 0.3711 & 0.3706 & $-$0.2752 & 1.2439 & 1.1847 & $-$0.9429 & 1.9583  \\
 $\Omega^i$(eval)   &  1.4069 & $-$1.2031 & 2.4063 & 0.6495 & $-$0.5841 & 2.4343 & 1.1480 & $-$1.1009 & 2.5217  \\
\end{tabular}
\vskip 1cm
\caption{Comparison of  ratios of determinants of adjacency matrices with ratios of theta functions for the critical Ising model in three different lattices: A, B and C.  Each lattice is characterized by the couplings and aspect ratio [$w_h $,$w_v $]($m_1$,$m_2$,k,$n_1$,$n_2$). The $L\rightarrow \infty$ ratios are compared with theta function ratios for a fitted period matrix $\Omega$(fit) and the period matrix  $\Omega$(eval) evaluated by the procedure of section \ref{section4} (first homology group basis H1).}
\label{table1}
\end{table}

\pagebreak

\begin{table} \centering
\begin{tabular}{|c|cccc|cccc|}
 \multicolumn{9}{c}{ first homology group basis: H1 }\\\hline
${\cal N}=1008$   &$\Omega^r_{11}$&$\Omega^r_{12}$&$\Omega^r_{21}$&$\Omega^r_{22}$&$\Omega^i_{11}$&$\Omega^i_{12}$&$\Omega^i_{21}$&$\Omega^i_{22}$  \\\hline
$\Omega$  & 0.370493 &  -0.275048 &  -0.275048 &   1.24336 &    0.643674 &  -0.566694  & -0.566694  &  2.38225  \\
$\Omega^*$& 0.370493 &  -0.275048 &  -0.275048  &  1.24336 &    0.655324  & -0.601642 &  -0.601642  &  2.48709  \\ \hline
${\cal N}=\infty$   &$\Omega^r_{11}$&$\Omega^r_{12}$&$\Omega^r_{21}$&$\Omega^r_{22}$&$\Omega^i_{11}$&$\Omega^i_{12}$&$\Omega^i_{21}$&$\Omega^i_{22}$  \\\hline
$\Omega$  & 0.370551 & -0.275225 & -0.275225 & 1.24389 &  0.648086 & -0.579928 & -0.579928 & 2.42195  \\
$\Omega^*$ & 0.370551 & -0.275225 & -0.275225 & 1.24389 &  0.650837 & -0.588183 & -0.588183 & 2.44671   \\\hline
  \multicolumn{9}{c}{  first homology group basis:    H2} \\\hline
  ${\cal N}=1008$  &$\Omega^r_{11}$&$\Omega^r_{12}$&$\Omega^r_{21}$&$\Omega^r_{22}$&$\Omega^i_{11}$&$\Omega^i_{12}$&$\Omega^i_{21}$&$\Omega^i_{22}$  \\\hline
$\Omega$   &  -0.871520 &  -0.216570 &  -0.216570 &  -0.220063  &   1.43105  &  0.334292  &  0.334292  &  0.396879  \\
$\Omega^*$  &   -0.871520 &  -0.216570 &  -0.216570 &  -0.220063  &   1.43385  &  0.328678  &  0.328678  &  0.408106  \\\hline
${\cal N}=\infty$    &$\Omega^r_{11}$&$\Omega^r_{12}$&$\Omega^r_{21}$&$\Omega^r_{22}$&$\Omega^i_{11}$&$\Omega^i_{12}$&$\Omega^i_{21}$&$\Omega^i_{22}$  \\\hline
$\Omega$ &   -0.871523 & -0.216564 & -0.216564 & -0.220075  &  1.43209  & 0.33221 &  0.33221 &  0.401042  \\
$\Omega^*$ &     -0.871523 & -0.216564 & -0.216564 & -0.220075  &  1.43275 & 0.330885 & 0.330885 & 0.403693  \\ \hline
\end{tabular}
\vskip 1cm
\caption{ The period matrix for the lattice B, evaluated for two choices (H1 and H2) for the basis of the first homology group. Results are shown for both the direct lattice period matrix $\Omega$ and the dual lattice period matrix $\Omega^*$ at finite size ${\cal N}=1008$ and in the thermodynamic limit extrapolation.}
\label{table2}
\end{table}

\pagebreak

\begin{table} \centering
\begin{tabular}{|c|c|cc|cc|cc|cc|}
\multicolumn{2}{|c|}{}& \multicolumn{4}{c|}{${\cal N}=1008$}& \multicolumn{4}{c|}{${\cal N}= \infty$} \\ \hline
\multicolumn{2}{|c|}{}& \multicolumn{2}{c|}{$\Omega$}& \multicolumn{2}{c|}{$(\Omega+\Omega^*)/2$}& \multicolumn{2}{c|}{$\Omega$}& \multicolumn{2}{c|}{$(\Omega+\Omega^*)/2$} \\ \hline

 H1  &    H2    &   H1  &    H2    &   H1   &    H2  &    H1 &     H2  &   H1 &   H2 \\ \hline
$\theta^4_1$&$\theta^4_1$ & 0.013898 & 0.013413 &  0.013402 & 0.013407 &  0.013525 & 0.013411 & 0.013409&  0.013409 \\
$\theta^4_6$&$\theta^4_{11}$ &0.352897&  0.372173 &  0.360822&  0.360606 &  0.358967&  0.363532 & 0.360852&  0.360852  \\
$\theta^4_7$&$\theta^4_{10}$ &0.000068&  0.000031 &  0.000049&  0.000050 &  0.000053&  0.000044&  0.000049&  0.000049  \\
$\theta^4_8$&$\theta^4_{12}$ &0.366408&  0.385027 &  0.373849&  0.373640 &  0.372113&  0.376527&  0.373884&  0.373884  \\
$\theta^4_{10}$&$\theta^4_7$ &1.000000&  1.000000 &  1.000000&  1.000000 &  1.000000&  1.000000&  1.000000&  1.000000 \\
$\theta^4_{11}$&$\theta^4_6$ &0.034803&  0.030549 &  0.032203&  0.032251 &  0.032809 & 0.031807&  0.032216&  0.032216 \\
$\theta^4_{12}$&$\theta^4_8$ &0.959525&  0.964607 &  0.962725&  0.962670 &  0.961978&  0.963155&  0.962706&  0.962707 \\
$\theta^4_{14}$&$\theta^4_{15}$ &0.811513&  0.812151 &  0.816897&  0.816887&   0.815469&  0.815670&  0.816764 & 0.816763  \\
$\theta^4_{15}$&$\theta^4_{14}$ &0.040773&  0.036186 &  0.037616&  0.037667 &  0.038358 & 0.037285&  0.037639 & 0.037639  \\
$\theta^4_{16}$&$\theta^4_{16}$ &0.777163&  0.782368 &  0.785101&  0.785040 &  0.783083 & 0.784349&  0.784960&  0.784959 \\ \hline
\end{tabular}
\vskip 1cm
\caption{Comparison of the theta function ratios for the period matrices of table \ref{table2}. The ratios corresponding to the period matrix evaluated with the H2 basis, are permuted to make modular invariance explicit.}
\label{table3}
\end{table}

\end{document}